\documentclass[prb,twocolumn,showpacs,floatfix]{revtex4}
\usepackage{graphics}
\usepackage{graphicx}
\usepackage{amssymb}
\begin{document}
\newcommand{\RR}{\mathrm{\mathbf{R}}}
\newcommand{\rr}{\mathrm{\mathbf{r}}}
\newcommand{\defin}{\stackrel{def}{=}}

\title{Effect of strain on the orbital and magnetic ordering of manganite thin films and their interface with an insulator}
\author{A. Baena}
\affiliation{Instituto de Ciencia de Materiales de Madrid (CSIC), Cantoblanco, 28049 Madrid (Spain)}
\author{L. Brey}
\affiliation{Instituto de Ciencia de Materiales de Madrid (CSIC), Cantoblanco, 28049 Madrid (Spain)}
\author{M.J. Calder\'on}
\affiliation{Instituto de Ciencia de Materiales de Madrid (CSIC), Cantoblanco, 28049 Madrid (Spain)}
\date{\today}

\begin{abstract}
We study the effect of uniform uniaxial strain on the ground state electronic configuration of a thin film manganite. 
Our model Hamiltonian includes the double-exchange, the Jahn-Teller electron-lattice coupling, and the antiferromagnetic superexchange. The strain arises due to the lattice mismatch between an insulating substrate and a manganite which produces a tetragonal distortion. This is included in the model via a modification of the hopping amplitude and the introduction of an energy splitting between the Mn e$_g$ levels. We analyze the bulk properties of half-doped manganites and the electronic reconstruction at the interface between a ferromagnetic and metallic manganite and the insulating substrate. The strain drives an orbital selection modifying the electronic properties and the magnetic ordering of manganites and their interfaces.  
\end{abstract}
\maketitle
\section{Introduction}
\label{sec:intro}
Manganites are strongly correlated oxides that show a large variety of magnetic and electronic phases due to a strong interrelation between the orbital, charge and spin degrees of freedom.~\cite{dagotto-book,israel07} They are particularly well known for the measured colossal magnetoresistance and the half-metallicity, which is responsible for the large tunneling magnetoresistance observed in manganite/insulator/manganite trilayers.~\cite{jo00,bibes-review} The current interest on oxide heterostructures~\cite{dagotto-perspectives07,huijben-AM09} and the electronic reconstruction occurring at their interfaces~\cite{ohtomo02,ohtomo04,okamoto-nat04} has also been reflected on a renewed interest on heterostructures involving manganites with different properties.~\cite{koida02,lin06,yamada06,brey-PRB07,smadici07,niebieskikwiat07,salafranca08,may08,linPRB08,calderonPRB08,nandaPRL08,yuPRB09,spin-filterAM10} In different all-manganite heterostructures, it has been observed that the properties of thin manganite layers may be modified with respect to their bulk behavior, for instance, by the appearance of a ferromagnetic moment in a nominally antiferromagnetic manganite~\cite{niebieskikwiat07,salafranca08,spin-filterAM10} or by the formation of a ferromagnetic two dimensional electron gas at the interface between two antiferromagnetic insulating manganites.~\cite{koida02,lin06,yamada06,smadici07,may08,linPRB08,calderonPRB08,nandaPRL08} Orbital reconstruction, a modification of the orbital occupancy at interfaces between different materials, has also been observed.~\cite{abadAFM07,chaklalian07,yuPRL10}

\begin{figure}
\includegraphics[clip,width=0.35 \textwidth]{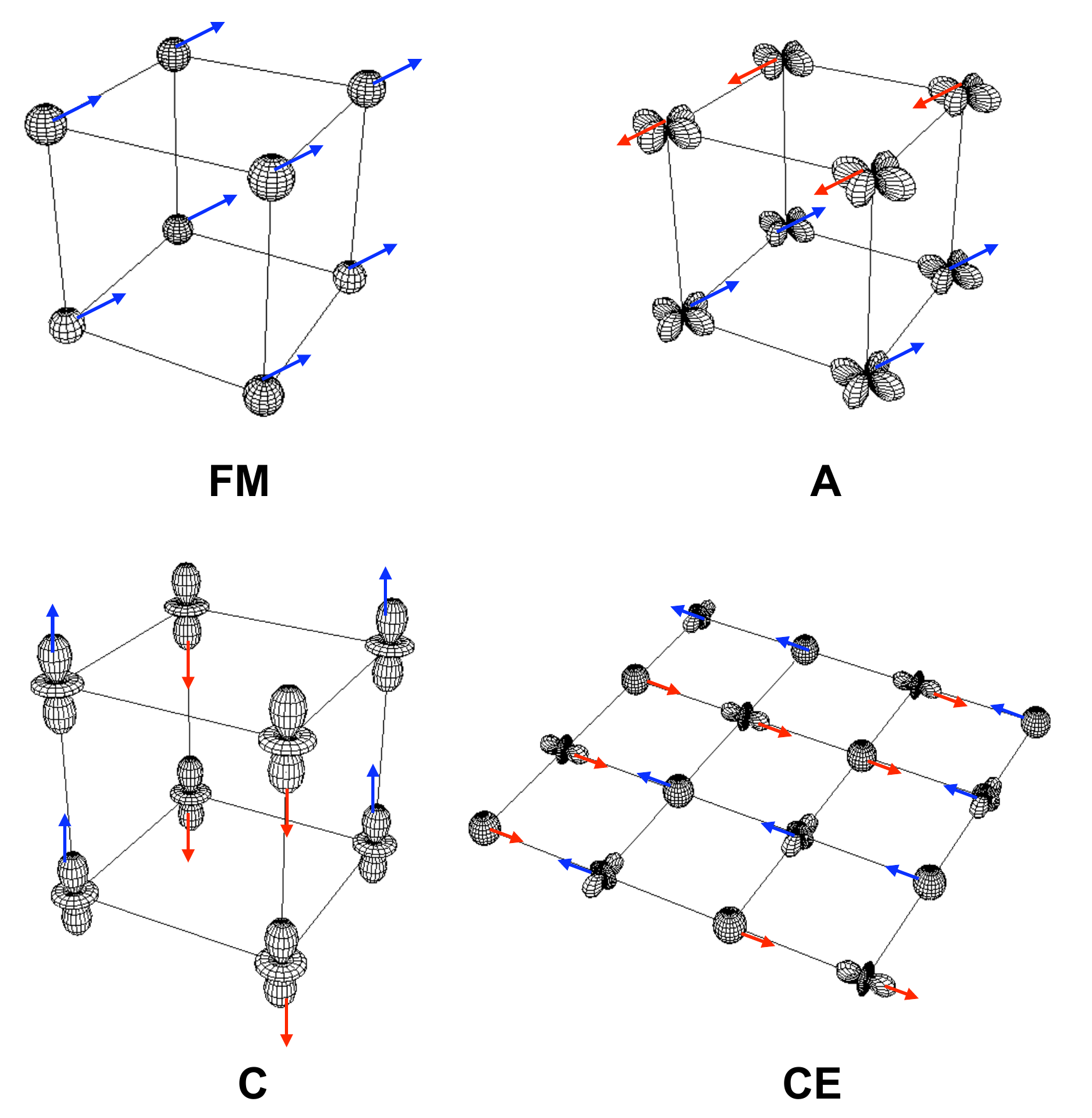}
\caption{(Color online) Cartoon of the possible spin and orbital order configurations for half-doped manganites. FM labels a three dimensional ferromagnet while the other three (A, C, and CE) are different antiferromagnetic orderings. The FM configuration is orbital disordered (represented here by isotropic spherical orbitals). The A-type, with FM planes coupled antiferromagnetically, favors the occupancy of the $x^2-y^2$ orbitals. The C-type, with FM lines in the z-direction coupled antiferromagnetically, favors the occupancy of the $3z^2-r^2$ orbitals. The CE-type ordering consists of FM zig-zag chains coupled antiferromagnetically (only the xy plane is shown here) and is associated with a peculiar orbital ordering (alternating $3x^2-r^2$, $3y^2-r^2$ and more isotropic orbitals) and checkerboard charge ordering. 
}
\label{fig:mag-conf}
\end{figure}

Manganites have the pseudocubic perovskite structure with chemical composition A$_{1-x}$A'$_x$MnO$_3$ with A typically a trivalent rare-earth (e.g. La or Pr) and A' a divalent cation (e.g. Ca or Sr). $1-x$ is the concentration of electrons moving on the Mn $e_g$ ($x^2-y^2$ and $3z^2-r^2$) orbital bands. Mn ions are in the center of oxygen octahedra that may undergo Jahn-Teller (JT) distortions. These lattice distortions couple to the charge and orbital degrees of freedom producing a splitting of the $e_g$ levels sometimes associated with charge and/or orbital ordering. The competition between the JT, the antiferromagnetic (AF) superexchange, and the kinetic energy via the double exchange (DE) interaction (Hund's coupling is assumed to be infinite), leads to a complex phase diagram as a function of composition and doping.~\cite{kajimoto02} In particular, at half-doping ($x=0.5$) many manganites are insulating and show charge, orbital and antiferromagnetic ordering (of the CE-type, see Fig.~\ref{fig:mag-conf}) while for $0.2 \lesssim x < 0.5$ a ferromagnetic (FM) metallic behavior is usually found. As illustrated in Fig.~\ref{fig:mag-conf} different magnetic orderings lead to diverse orbital configurations. The antiferromagnetic configurations are also insulating while ferromagnetism is usually accompanied by metallicity by virtue of the DE interaction.

Manganite multilayers and thin films are epitaxially grown on insulating substrates. The lattice mismatch between the different layers gives rise to a uniaxial strain which may affect the bulk properties of a thin manganite~\cite{fang00,ogimotoAPL01,infante07,adamoAPL09,sadocPRL10,nanda09} and/or the electronic reconstruction at its interface.~\cite{yamada06} Strain may also be responsible for phase separation.~\cite{ahn04,sagdeoJAP08}
The inplane strain may range from $\sim -2.3\%$ to $\sim 3.2\%$ depending on the substrate and the growth direction.~\cite{adamoAPL09} This strain modifies the relation between the lattice parameter in the direction perpendicular to growth ($c$) and the one in the parallel plane ($a$).

The possible modifications produced by strain in a manganite are twofold:~\cite{millisJAP98} (i) the reduction (increase) of the lattice parameter in a particular direction would lead to an increase (reduction) of the hopping amplitude and, (ii) a distortion of the pseudocubic symmetry leads to a splitting of the $e_g$ levels which may produce orbital ordering via the Jahn-Teller coupling.~\cite{tokura00,abadAFM07,nandaPRB08,sadocPRL10,dongPRB10} These mechanisms alter the competition between the localizing and delocalizing interactions in manganites in opposite directions. For instance, in a (001) thin film, a compressive strain (reduction of the lattice parameter in the xy plane) would increase the hopping within the xy plane and decrease it in the z-direction favoring the $x^2-y^2$ e$_g$ orbitals (rather than the $3z^2-r^2$ orbitals) to be occupied. On the other hand, a compressive tetragonal distortion would produce a lowering of the $3z^2-r^2$ orbitals with respect to the $x^2-y^2$ ones.~\cite{tokura00,abadAFM07,nandaPRB08,sadocPRL10} Experimentally,~\cite{tokura00} it is observed that when the lattice parameter in the z-direction is larger than in the xy plane (namely, $c/a>1$) a C-type AF ordering with occupied $3z^2-r^2$ orbitals is favored in contrast to the A-type AF configuration with occupied $x^2-y^2$ orbitals which occurs when $c/a<1$. These observations are consistent with a predominance of mechanism (ii) over (i). 

Here we analyze the effect of strain on homogeneously strained epitaxially grown manganites in the $(001)$ direction. In Sec.~\ref{sec:bulk} we study the modifications produced by strain in the phase diagram of half-doped ($x=0.5$) bulk manganites. In Sec.~\ref{sec:interface} we turn the focus to the interface layers between a ferromagnetic metallic manganite with $x=0.3$ and the insulating substrate and study how the electronic reconstruction is affected by strain. We finish in Sec.~\ref{sec:disc-concl} with a discussion on the light of reported experimental results and the conclusions.  

\section{Effect of strain in bulk manganites at half-doping}
\label{sec:bulk}
\subsection{Model}
\label{subsec:bulk-model}

\begin{figure*}
\includegraphics[clip,width=0.9 \textwidth]{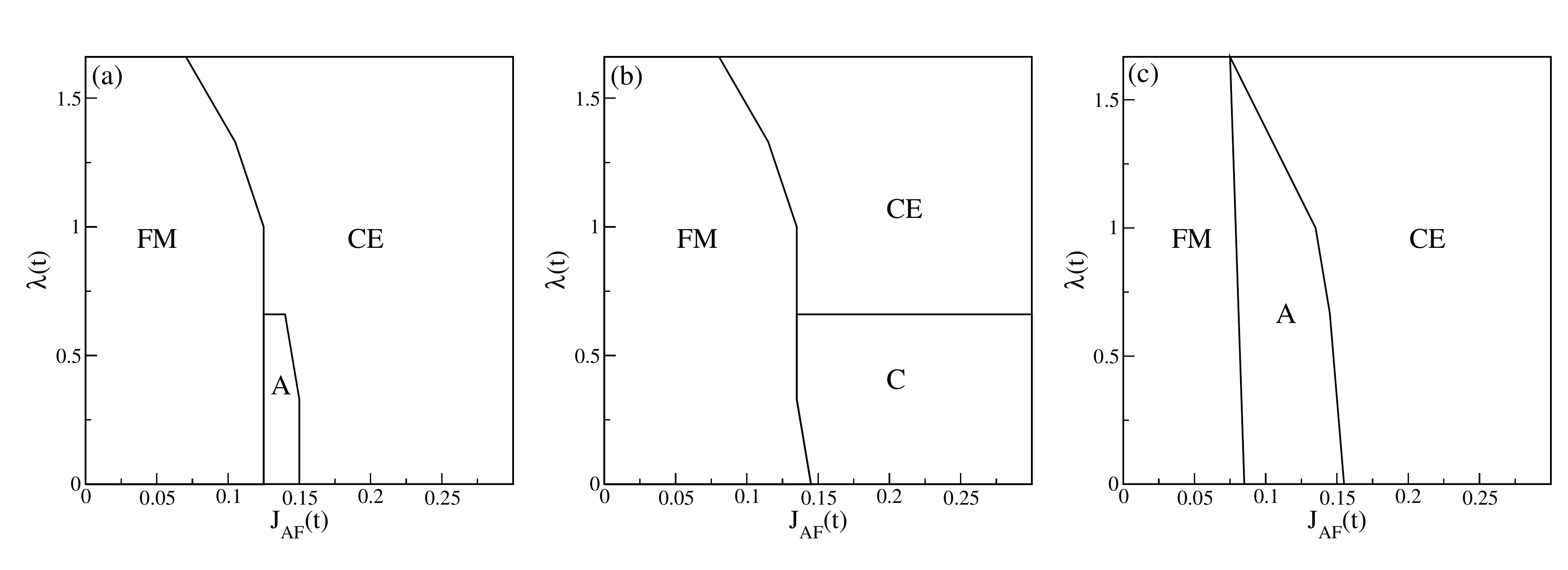}
\caption{ Bulk phase diagram $\lambda$ versus $J_{AF}$ for half-doped manganites. FM, A, C and CE label the different magnetic and orbital orders considered (see Fig.~\ref{fig:mag-conf}). (a) Without strain. (b) With compressive strain $e_{xy}=-2\%$. (c) With tensile strain $e_{xy}=2\%$. The splitting between the e$_g$ levels is $\delta=50 e_{xy} t$ [namely, $|\delta|=t$ in (b) and (c)]. 
}
\label{fig:bulk-lambda-JAF}
\end{figure*} 
\begin{figure*}
\includegraphics[clip,width=0.6 \textwidth]{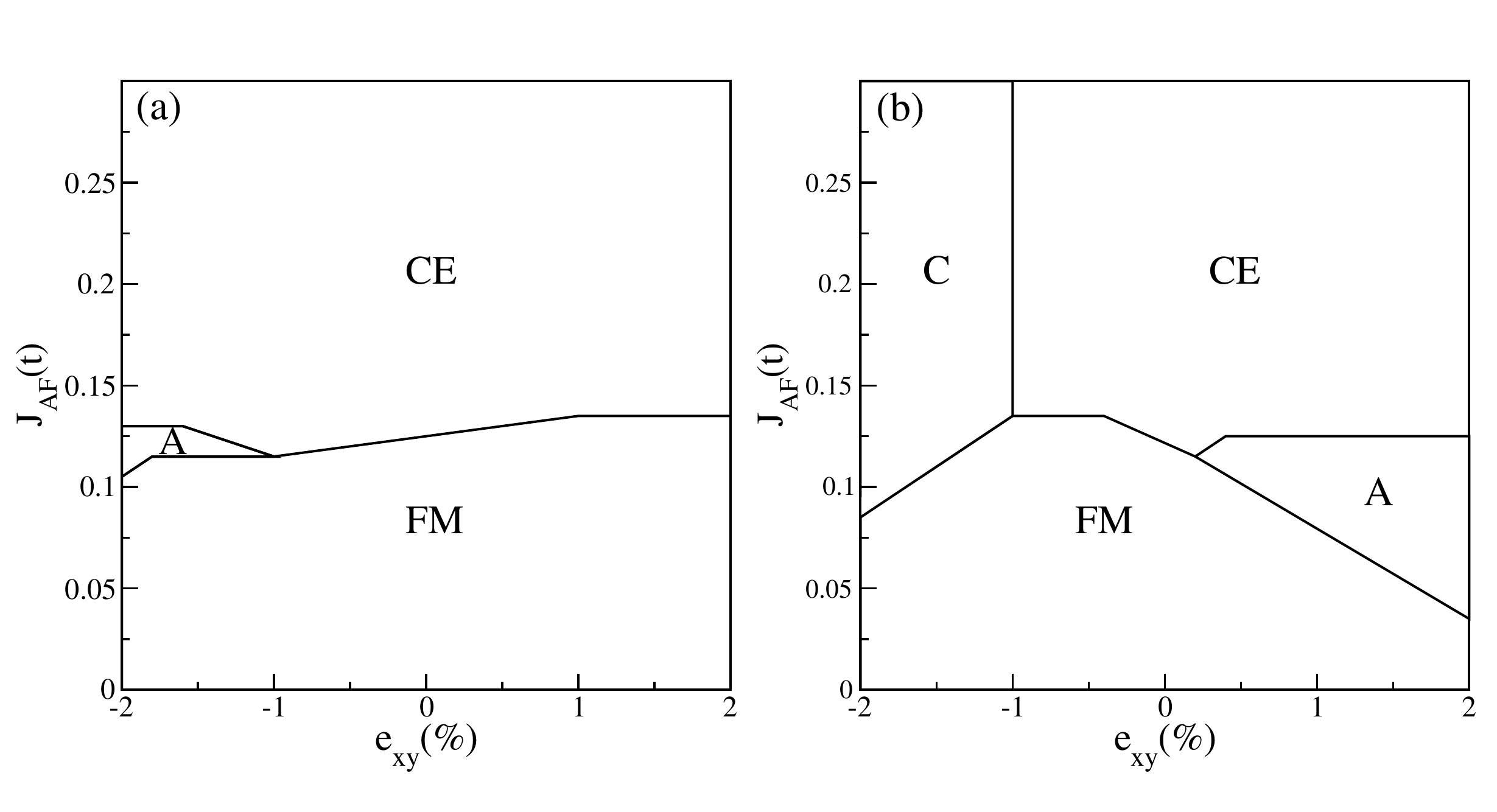}
\caption{ Bulk phase diagrams $J_{AF}$ versus $e_{xy}$ for $\lambda=1 t$ with $\delta=0$ (a) and $\delta=100 e_{xy} t$ (b).
}
\label{fig:bulk-JAF-exy}
\end{figure*}   

In the absence of strain, the model Hamiltonian for manganites includes the kinetic energy, the electron-lattice Jahn-Teller coupling, and the nearest neighbor antiferromagnetic superexchange~\cite{brey-PRL04}  

\begin{eqnarray}
H &=&- \sum_{i,j,\gamma,\gamma'} f_{i,j} t^u_{\gamma,\gamma'} C_{i,\gamma}^{\dagger} C_{j,\gamma'} \nonumber \\
&+& \sum_i \left[\beta Q_{1i}^2+Q_{2i}^2+Q_{3i}^2
+ \lambda \left(Q_{1i} \rho_i+ Q_{2i} \tau_i^x+Q_{3i} \tau_i^z \right) \right] \nonumber \\
&+&\sum_{i,j}  J_{\rm AF}^{ij} {\mathbf S}_i  {\mathbf S}_j \,
\label{eq:H}
\end{eqnarray}
where $C_{i,\gamma}^{\dagger}$ creates an electron on the Mn i-site in the $e_g$ orbital $\gamma$ ($\gamma=1,2$ with $1=|x^2-y^2 \rangle$ and $2=|3 z^2-r^2 \rangle$). The hopping amplitude $f_{i,j}$ depends on the Mn core spins orientation given by the spherical angles $\theta$ and $\psi$ via  the double-exchange mechanism 
{\small
\begin{equation}
f_{i,j}=\cos(\theta_i/2) \cos(\theta_j/2)+ \exp[i(\psi_i-\psi_j)]\sin(\theta_i/2) \sin(\theta_j/2)  \, , 
\end{equation}}
and on the orbitals involved $t^{\rm x(y)}_{1,1}=\pm \sqrt{3} \,t^{\rm x(y)}_{1,2}=\pm \sqrt{3}\, t^{\rm x(y)}_{2,1}=3\, t^{\rm x(y)}_{2,2}=t_o^{x,y}$, and $t^{\rm z}_{2,2}=t_o^z$ where the superindices x,y, and z refer to the direction in the lattice.  In the absence of strain $t_o^{x,y}=t$ and $t_o^z=4/3 t$. All the parameters are given in units of $t$ which is estimated to be $\sim 0.1 - 0.5$ eV depending on the chemical composition.~\cite{review_ramirez} $Q_{li}$, with $l=1,2,3$, are the phonon modes of the oxygen octahedra. $\rho_i=\sum_{\gamma} C_{i,\gamma}^{\dagger}C_{i,\gamma}$ is the site occupation. The orbital occupation is represented by the Pauli matrices $\tau^{x,z}$: $\tau_i^x=2 Re ( C^{\dagger}_{i,3z^2-r^2}C_{i,x^2-y^2})$ and $\tau_i^z=C^{\dagger}_{i,x^2-y^2}C_{i,x^2-y^2}-C^{\dagger}_{i,3z^2-r^2}C_{i,3z^2-r^2}$. $\beta$ is assumed to be very large so the $Q_1$ breathing mode is frozen. $\lambda$ is the Jahn-Teller coupling and $J_{\rm AF}$ is the antiferromagnetic superexchange coupling between the localized t$_{2g}$ spins which is estimated to be $\sim 1-10$ meV~[\onlinecite{dagotto-book}].

The strain is introduced uniformly in a cubic system ($c/a=1$). We consider uniaxial strain arising from the lattice mismatch between the manganite and a cubic substrate and assume a $(001)$ growth direction. The strain can be tensile  (extension in the xy plane and compression in the z direction: $c/a<1$) or compressive (compression in the xy plane and extension in the z direction: $c/a>1$). The strain in the xy-plane $e_{xy}$ is defined as $e_{xy}=(a_s-a)/a_s$ with $a$ the average lattice parameter of the manganite and $a_s$ the in-plane lattice parameter of the substrate. Therefore, $e_{xy}>0$ ($<0$) corresponds to tensile (compressive) strain. The relation between the strain in the z-direction $e_z$ and $e_{xy}$ is given by the Poisson ratio $\nu$ as $e_z=-4 \nu e_{xy}$ with $0.3 \lesssim  \nu \lesssim 0.4$ for manganites.~\cite{yamada06,nandaPRB08,adamoAPL09} We choose $e_z=-{\frac{3}{2}} e_{xy}$~[\onlinecite{ahnPRB01}] and allow $e_{xy}$ to range between $-0.02$ ($-2\%$) and $0.02$ ($2\%$). 

The effect of the strain on the system is twofold. On one hand, it affects the overlapping matrices and, therefore the hopping amplitudes as~\cite{harrisonbook2}
\begin{eqnarray}
&t_o^{x,y}& = t(1-2 e_{xy}) \nonumber \\
&t_o^z & = 4/3 t (1-2 e_z) \, .
\label{eq:hop-strain}
\end{eqnarray}
The strain can also induce a splitting $\delta$ of the $e_g$ orbitals:~\cite{millisJAP98,tokura00,nandaPRB08,sadocPRL10} tensile strain lowers the energy of the $|x^2-y^2 \rangle$ orbital with respect to $|3z^2-r^2 \rangle$ ($\tau_z >0$) while compressive strain does the opposite ($\tau_z <0$). This is introduced as an extra term in the Hamiltonian 
\begin{equation}
H_{\delta}= \sum_{\gamma} \epsilon_{\gamma} \sum_{i} C_{i,\gamma}^{\dagger} C_{i,\gamma} \, ,
\label{eq:delta}
\end{equation}   
with $\epsilon_{3z^2-r^2}=\delta/2$ and $\epsilon_{x^2-y^2}=-\delta/2$. We have analyzed a range of values $|\delta| \le 100 |e_{xy}| t$ (namely $|\delta| \le 2t$).~\cite{nandaPRB08} 

We find the ground state configuration of a half-doped manganite by solving the Hamiltonian in Eq.~[\ref{eq:H}] plus the term in Eq.~[\ref{eq:delta}] self-consistently (at zero temperature) in a $4 \times 4 \times 4$ system with periodic boundary conditions in the three directions. The phase diagrams as a function of $\lambda$, $J_{\rm AF}$ and $e_{xy}$  result from comparing the energies of the different configurations illustrated in Fig.~\ref{fig:mag-conf}.

\subsection{Results}
\label{subsec:bulk-results}
The phase diagram as a function of $\lambda$ and $J_{AF}$ for a half-doped manganite is well known~\cite{breyx05PRB05} and is shown in Fig.~\ref{fig:bulk-lambda-JAF}(a) as a reference. For small values of $J_{AF}$ and $\lambda < 1.6 t$, the ground state is ferromagnetic and metallic. For $\lambda > 1.6 t$ (not shown), the FM phase is insulating. As $J_{AF}$ increases, and for $\lambda \lesssim 0.7 t$, two different antiferromagnetic phases arise: (i) For a narrow range of values of $J_{AF}$, an A-type AF phase consisting of FM xy-planes coupled antiferromagnetically in the z-direction, and (ii), for the largest realistic values of $J_{AF}$, the CE AF order. The CE order consists of FM zig-zag chains coupled AF between them. As $\lambda$ increases, this CE phase becomes dominant. 

In Fig.~\ref{fig:bulk-JAF-exy} the effect of the strain on the ground state configuration is shown for $\lambda=1t$. In Fig.~\ref{fig:bulk-JAF-exy}(a), the splitting $\delta$ between the e$_g$ levels caused by the strain is neglected so the only effect of the strain is to modify the hoppings as described in Eq.~[\ref{eq:hop-strain}]. Compressive strain ($e_{xy}<0$) produces a decrease of the hopping in the z-direction and an enhancement in the xy-plane causing the FM configuration to lose energy with respect to the A-AF configuration. On the contrary, tensile strain increases the hopping in the z-direction and lowers it in the xy-plane. In this case, both AF phases (A and CE), with zero hopping in the z-direction due to the antiferromagnetic order, lose kinetic energy with respect to the FM phase. 

When the splitting $\delta$ is included (Eq.~\ref{eq:delta}), the changes in the phase diagram are more dramatic, see Fig.~\ref{fig:bulk-JAF-exy}(b). As illustrated in Fig.~\ref{fig:mag-conf}, the different AF configurations are related to specific orbital orderings. In terms of the pseudospin $\tau_z$, the C-AF ordering has $\tau_z<0$ (preferred occupation of the $3z^2-r^2$ orbital) while both the A and CE orderings have $\tau_z >0$ (preferred occupation of the $x^2-y^2$ orbital). The term in Eq.~\ref{eq:delta} tends to enforce a particular value of $\tau_z$. Therefore, the A phase becomes the ground state for tensile strain ($\delta >0$) due to the lowering of the $x^2-y^2$ orbital with respect to the $3z^2-r^2$ and a C phase dominates for compressive strain and $J_{AF} \gtrsim 0.1 t$. This C phase appears for $|\delta| > 50 |e_{xy}| t$ if $\lambda=0$ and for $|\delta| > 60 |e_{xy}| t$ if $\lambda=1 t$. Comparing Figs.~\ref{fig:bulk-JAF-exy}(a) and (b), it is apparent that the effect of the strain on the splitting of the e$_g$ levels clearly overcomes the effect caused by the modifications in the hopping for a sufficiently large $\delta \gtrsim 50 |e_{xy}| t$.

In Fig.~\ref{fig:bulk-lambda-JAF}~(b) and (c) the modifications on the $\lambda$ versus $J_{AF}$ phase diagram caused by compressive and tensile $2\%$ strain are illustrated with $|\delta|=t$. Consistently with the results in Fig.~\ref{fig:bulk-JAF-exy}(b), the C phase arises when compressive strain is applied, while the A phase becomes more prominent with tensile strain. It is also noticeable in Fig.~\ref{fig:bulk-lambda-JAF}~(b) that the C ordering is not favored by the JT coupling $\lambda$.

\section{Effect of strain at a manganite/insulator interface}
\label{sec:interface}
\subsection{Model}
\label{subsec:interface-model}
\begin{figure}
\includegraphics[clip,width=0.45 \textwidth]{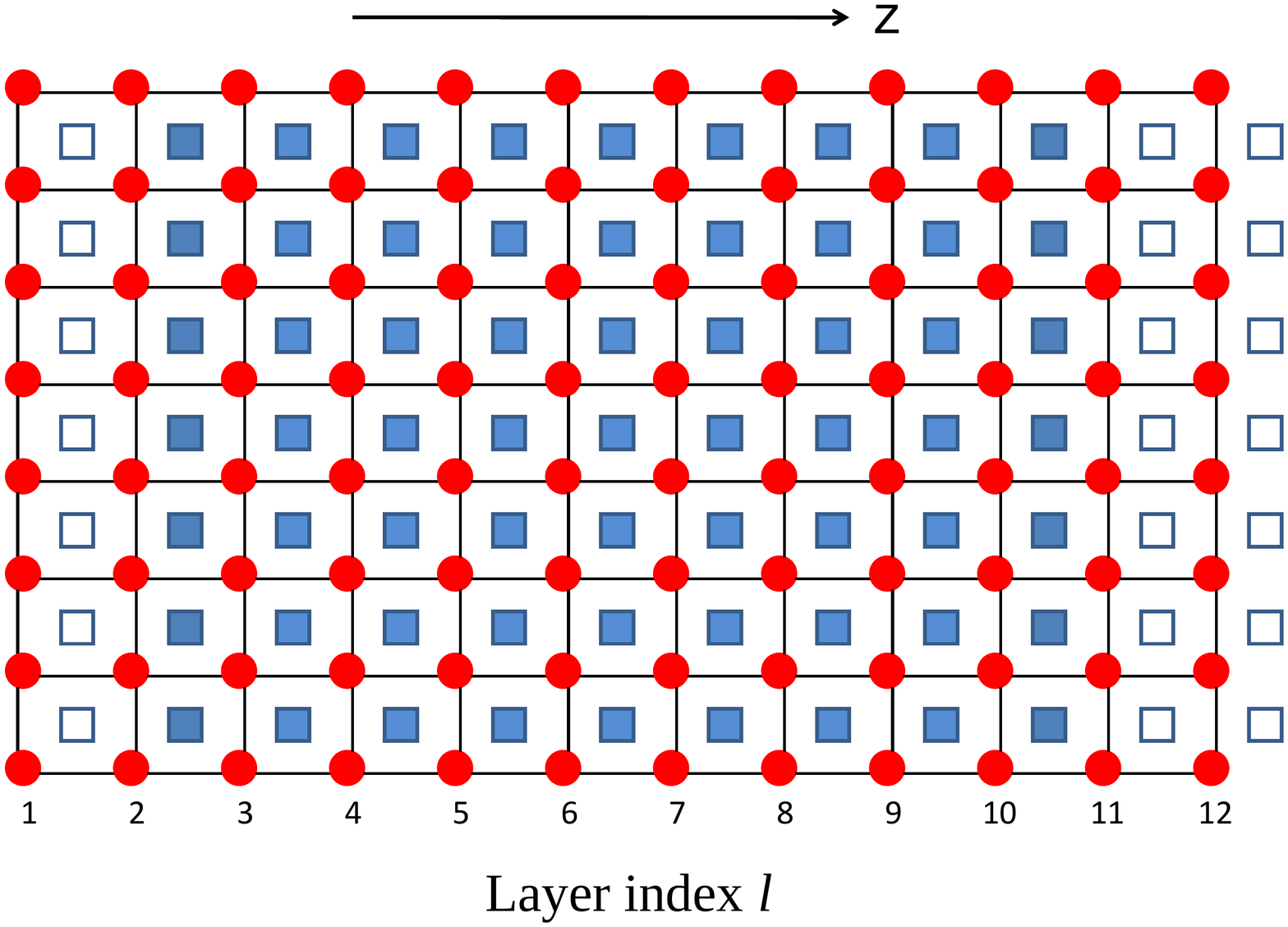}
\caption{ (Color online) 2-dimensional (xz plane) projection
of the considered heterostructure. The circles represent the Mn sites while the squares are the A-sites (La, Sr, Ca, etc) shifted by $(1,1,1) {{a}\over{2}}$ with respect to the Mn. $a$ is the lattice parameter.
Full squares represent the A$^{3+}_{0.7}$A'$^{2+}_{0.3}$O$^{2-}$ plane, with a charge
density of $+0.7$ per A atom. Empty squares correspond to the
Sr$^{2+}$O$^{2-}$ planes where the charge density is zero. Periodic boundary conditions are applied in all three directions.  
}
\label{fig:mang-ins}
\end{figure}

We turn now to analyzing the effect of the strain on the electronic reconstruction at a (001) manganite/insulator interface. We focus on the case of a manganite with $x=0.3$, which corresponds to a ferromagnetic and metallic bulk ground state. 
In the $(001)$ direction, manganites alternate MnO$_2$ and AO planes as shown in Fig.~\ref{fig:mang-ins}. The AO planes give a positive background charge $1-x$. 
We consider a thin manganite slab with a $4 \times 4$ cross section and $l_z=12$ Mn planes in the z-direction. The insulator is included as a hard-wall fixing the boundary condition at $l=1$ and $l=l_z$ to zero charge density.~\cite{brey-PRB07} The interface layers are $l=2$ and $l=l_z-1$.

The model Hamiltonian is the same as described for the bulk case in Sec.~\ref{subsec:bulk-model} plus a Hartree term that takes into account the long range Coulomb interaction between all the charges in the system.~\cite{brey-PRB07,salafranca08,calderonPRB08} $H_{\rm Hartree}$ takes the form
{\small
\begin{equation}
H_{\rm Hartree} ={{\frac{e^2}{\epsilon}}} \sum_{i \ne j} \left({\frac{1}{2}} {\frac {\langle n_i \rangle   \langle n_j \rangle}{|{\mathbf R}_i-{\mathbf R}_j |}} +{\frac{1}{2}} {\frac {Z_i Z_j}{|{\mathbf R}^A_i-{\mathbf R}^A_j |}}
-{\frac{Z_i   \langle n_j \rangle}{|{\mathbf R}^A_i-{\mathbf R}_j |}}\right)
\label{eq:hartree}
\end{equation}
}%
with ${\mathbf R}_i$ the position of the Mn ions, $\langle n_i \rangle=\sum_{\gamma} \langle  C_{i,\gamma}^{\dagger} C_{i,\gamma} \rangle$ the occupation number on the Mn i-site, $eZ_i$ the charge of the A-cation located at ${\mathbf R}_i^A$, and $\epsilon$ the dielectric constant of the material. The relative strength of the Coulomb interaction is given by the parameter $\alpha=e^2/a \epsilon t \sim 1-2$~[\onlinecite{lin06}], where $\epsilon$ is the dielectric constant of the manganite. 

The charge density is $\sim 0.7$ in the central planes where the bulk values for the $x=0.3$ manganite are recovered. However, close to the insulator, the charge density decreases towards $0$ to fulfill the boundary condition. A redistribution of charge occurs to screen the positive charge background and is controlled by the Coulomb parameter $\alpha$. Experimentally, the charge transfer between different layers occurs within $2-3$ unit cells, ~\cite{koida02} where the charge density may be close to half-doping and the CE-type AF ordering.

We consider different possible configurations at the manganite-insulator interface and compare their energies to define the phase diagrams as a function of $\lambda$, $J_{AF}$, $e_{xy}$, and $\alpha$:~\cite{brey-PRB07} (i) FM corresponds to all ferromagnetic planes; (ii) 1CE stands for a configuration with all ferromagnetic planes except for a single CE plane at the manganite-insulator interfaces; and (iii) 2CE includes two CE planes at the manganite-insulator interfaces.

\subsection{Results}
\label{subsec:interface-results}
\begin{figure}
\includegraphics[clip,width=0.45 \textwidth]{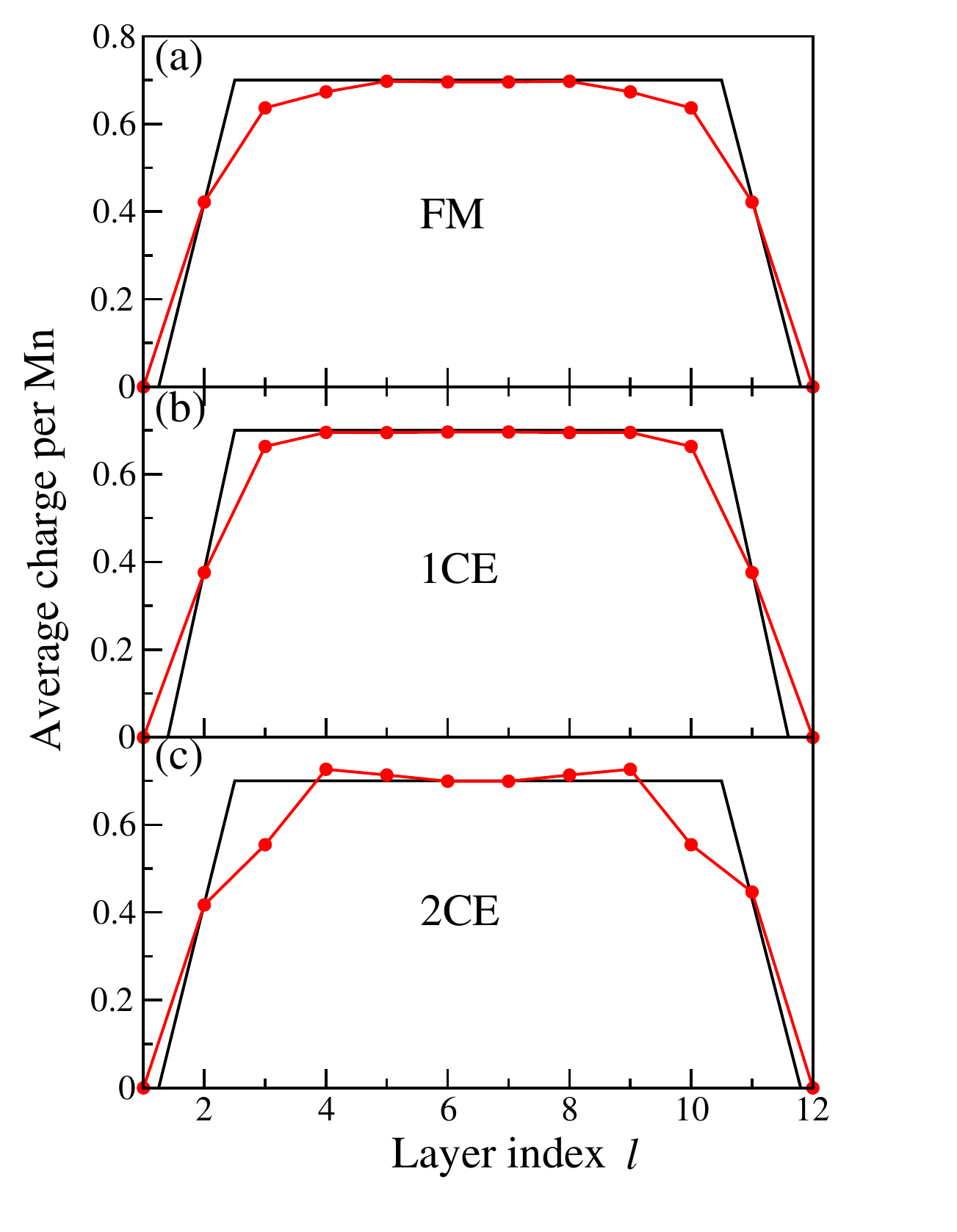}
\caption{(Color online) Average charge per plane for the three possible configurations considered at a manganite/insulator interface for $\lambda=1$, $\alpha=1$ and $e_{xy}=0$. The straight lines in black represent the positive background charge. (a) FM: all ferromagnetic planes, (b) 1CE: all ferromagnetic planes except for a single CE plane at each manganite/insulator interface (layers $l=2,11$), and (c) 2CE: two CE planes at each manganite insulator interface (layers $l=2,3,10,11$). 
}
\label{fig:cpp}
\end{figure}
\begin{figure}
\includegraphics[clip,width=0.34 \textwidth]{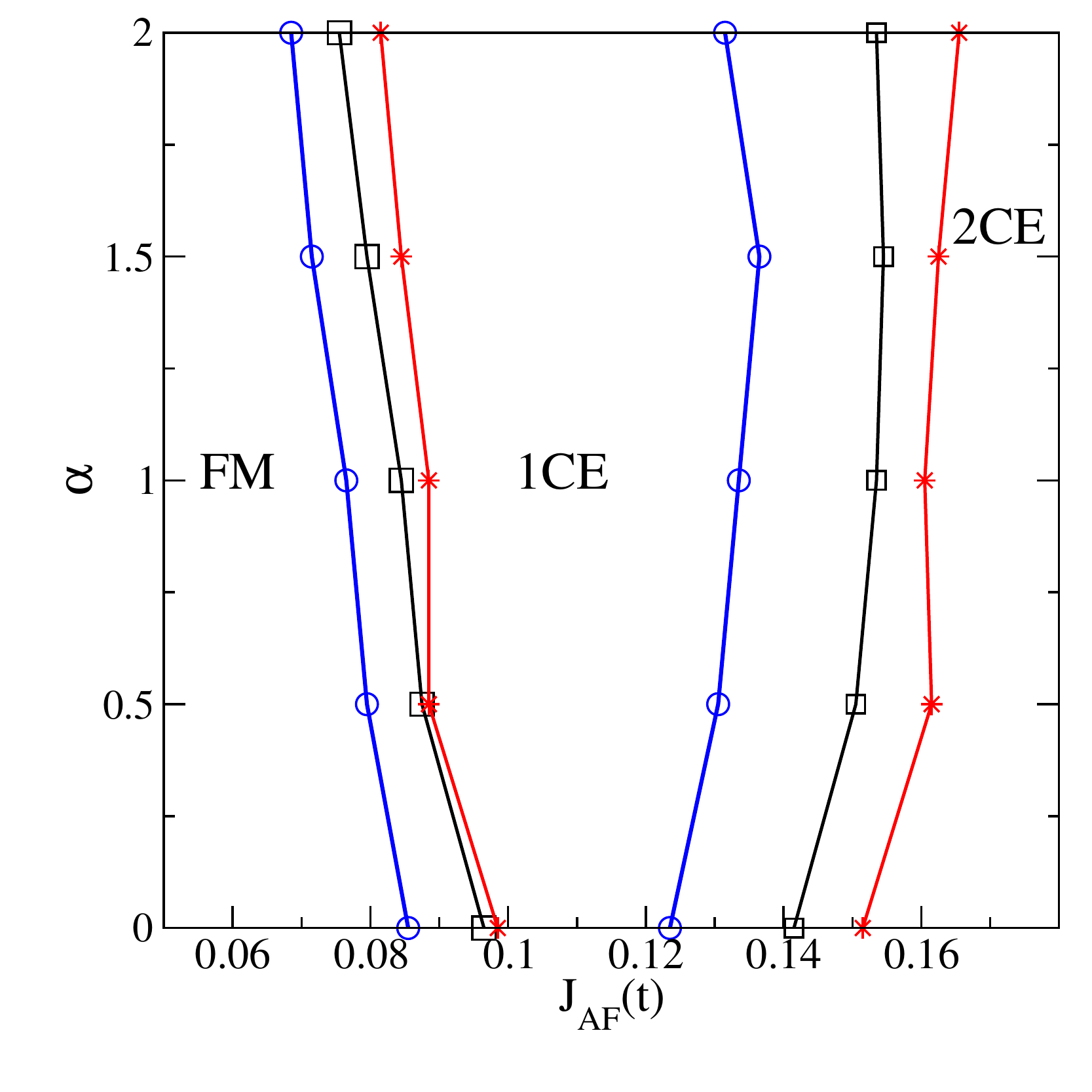}
\caption{ (Color online) Phase diagram $\alpha$ versus $J_{AF}$ for the manganite/insulator interface with $\lambda=1$. FM stands for all FM Mn planes, 1CE stands for a configuration with a single CE layer at the manganite surface, and 2CE stands for two CE layers at the manganite surface. The lines represent the boundaries between the different ground state configurations: The black lines (squares) correspond to the results with $e_{xy}=0$, the red lines (stars) are for compressive strain $e_{xy}=-2\%$, and the blue lines (circles) for tensile strain $e_{xy}=2\%$. $\delta=50 e_{xy} t$ is assumed.
}
\label{fig:ecoul-JAF}
\end{figure}
\begin{figure}
\includegraphics[clip,width=0.34 \textwidth]{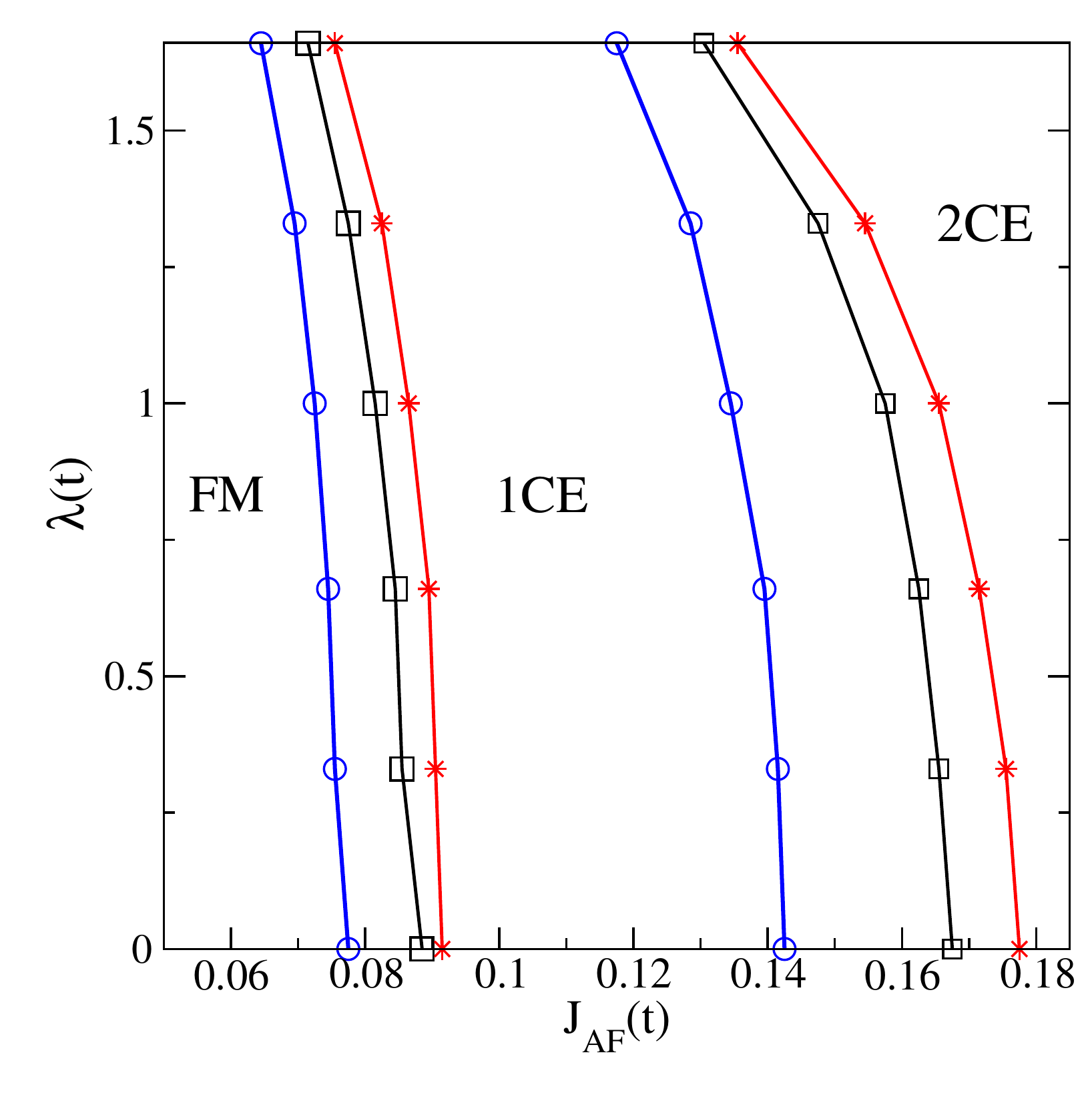}
\caption{(Color online) Phase diagram $\lambda$ versus $J_{AF}$ for the manganite/insulator interface with $\alpha=1$. The labels are the same as in Fig.~\ref{fig:ecoul-JAF}. $\delta=50 e_{xy} t$ is assumed.
}
\label{fig:lambda-JAF-mang-ins}
\end{figure}

In Fig.~\ref{fig:cpp} the average charge per plane in an insulator/manganite/insulator trilayer is shown for the three possible ground state configurations at the interface for $\lambda=1t$, $\alpha=1$ and $e_{xy}=0$. The redistribution of charge depends only very weakly on the value of the strain (not shown). The CE order opens a gap in the density of states at $x=0.5$ so, in the case of $\alpha=0$ it would tend to pin the charge density to this value.~\cite{brey-PRB07} However, for a finite $\alpha$, the Coulomb term Eq.~\ref{eq:hartree} controls the charge redistribution in such a way that the charge density is not pinned at $0.5$ in the CE layers and it is controlled instead by the screening of the positive background charge by means of the Hartree term in  Eq.~\ref{eq:hartree}.

Fig.~\ref{fig:ecoul-JAF} shows the $\alpha$ versus $J_{AF}$ phase diagram for three different values of the strain: $0$, $2\%$ and $-2\%$ with $\lambda=1t$ and $\delta=50 e_{xy} t$. This phase diagram without strain was studied before in Ref.~[\onlinecite{brey-PRB07}] with a model which neglected Jahn-Teller interactions but considered instead an interorbital Hubbard term to stabilize the antiferromagnetic phases. Those results compare very well to our $e_{xy}=0$ results in Fig.~\ref{fig:ecoul-JAF} (squares). A compressive ($e_{xy}<0$) strain makes the FM configuration relatively more stable (namely, a larger $J_{AF}$ is required to produce an antiferromagnetic CE order in one or two layers) while a tensile ($e_{xy}>0$) strain lowers the energy of the antiferromagnetic CE interface configurations with respect to the FM. These results are consistent with the behavior observed for bulk manganites in Figs.~\ref{fig:bulk-lambda-JAF} and \ref{fig:bulk-JAF-exy} (b) and are due to the preferred occupation of the $x^2-y^2$ orbitals in the CE order.

\begin{figure*}
\includegraphics[clip,width=0.6 \textwidth]{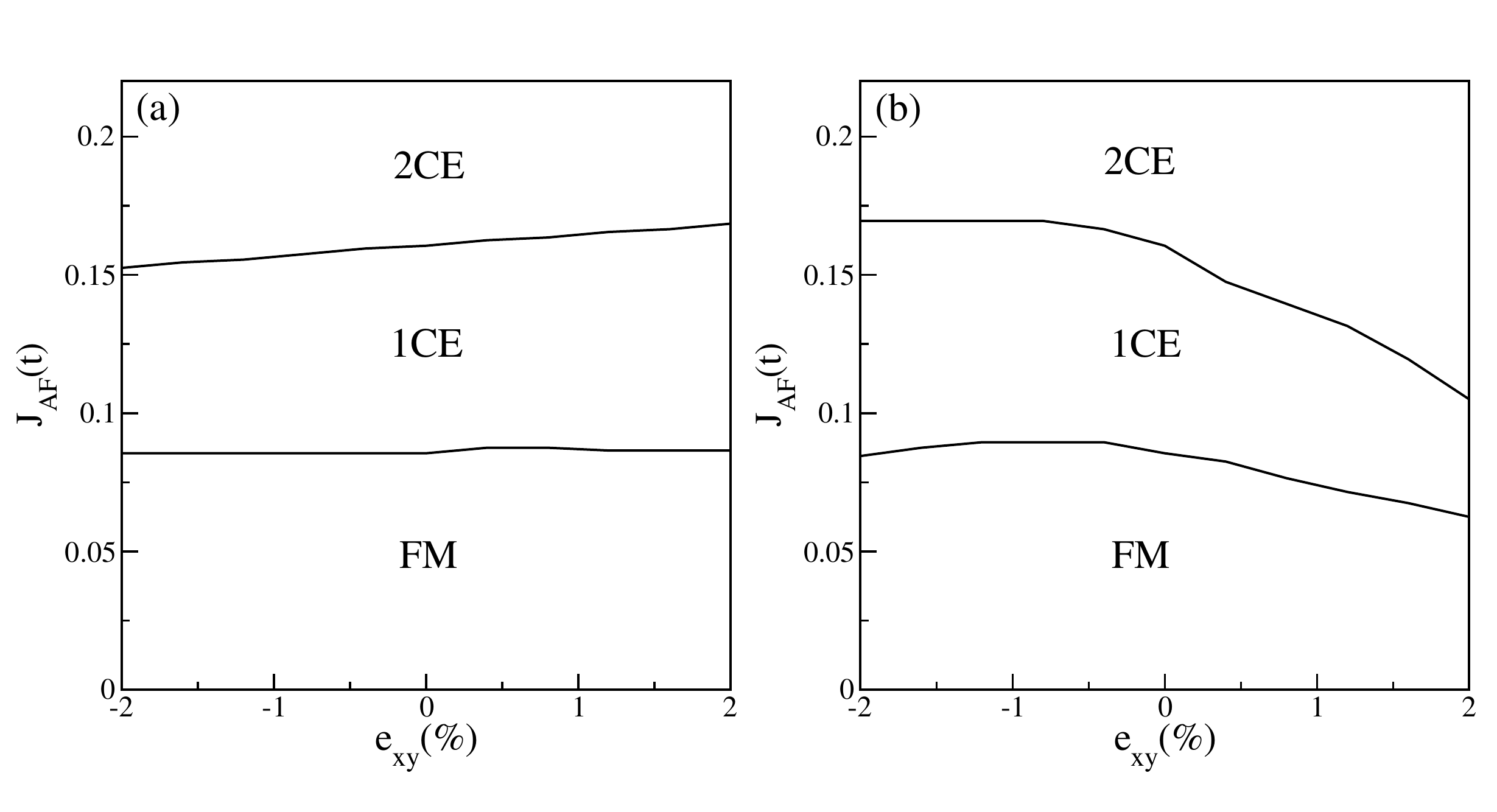}
\caption{Phase diagrams $J_{AF}$ versus $e_{xy}$ for the manganite/insulator interface with $\lambda=1$ and $\alpha=1$. The strain induced $e_g$ level splitting is $\delta=0$ in (a) and $\delta=100 e_{xy} t$ in (b).   
}
\label{fig:JAF-exy-mang-ins}
\end{figure*}

Similar phenomenology is observed in Fig.~\ref{fig:lambda-JAF-mang-ins} where the $\lambda$ versus $J_{AF}$ phase diagram is plotted. In general terms, tensile strain tends to favor the antiferromagnetic CE order close to the insulator while compressive strain does not seem to affect the interface ground state very strongly. Finally, we show the $J_{AF}$ versus $e_{xy}$ phase diagram in Fig.~\ref{fig:JAF-exy-mang-ins} for $\lambda=1t$ and $\alpha=1$. In Fig.~\ref{fig:JAF-exy-mang-ins} (a) $\delta=0$ so the only effect of the strain is to modify the hopping amplitude as given in Eq.~\ref{eq:hop-strain}. The effect of the strain is very mild in this case with only a slight gain of the configurations with CE planes for compressive ($e_{xy}<0$) strain. The tendency is the opposite and the dependence on strain is stronger when the splitting $\delta$ of the e$_g$ levels is included, as illustrated in Fig.~\ref{fig:JAF-exy-mang-ins}(b) for $\delta=100 e_{xy} t$.

At interfaces orbital reconstruction may arise.~\cite{calderonsurf,abadAFM07,chaklalian07,yuPRL10,okamotoPRB10} In particular, due to the breaking of the translational symmetry in the z-direction, the $3z^2-r^2$ orbital cannot gain much kinetic energy producing a splitting of the e$_g$ levels which favors the occupation of the $x^2-y^2$ orbital. In the all FM layers case, in the absence of strain, we observe this kind of ferro-orbital configuration (which corresponds to $\tau^z>0$) at the interface layer. A positive strain ($e_{xy}>0$) enhances this ferro-orbital interface ordering which also occurs in the bulk, as discussed in Sec.~\ref{sec:bulk}. On the other hand, a negative strain would produce an e$_g$ splitting opposite to the preferred one at an interface, leading for a sufficiently large value of the strain to a reduction and, eventually, to a sign change of $\tau^z$. Therefore, $\tau^z$ or, equivalently, the $Q_3$ phonon mode, increases as the strain goes from compressive to tensile. $\tau^x$ is nonzero only for the CE phase and has opposite signs on the sites with $3x^2-r^2$ and $3y^2-r^2$-like orbitals (see Fig.~\ref{fig:mag-conf}). $|\tau^x|$ is a measure of the mixing of the two e$_g$ levels and therefore decreases as $\tau_z$ increases. 

CE layers adjacent to a FM layer, as we have in both the 1CE and the 2CE configurations, show a charge and orbital distribution which is different from the isolated case. In particular, half of the spins are parallel to the spins in the nearby FM layer which produces a larger occupation on those sites (both on the CE and the FM layers) with respect to the sites with antiparallel spins.~\cite{calderonPRB08} Moreover, the redistribution of charge at the interface controlled by the Hartree term leads, in general, to a charge density away from the $0.5$ which stabilizes CE (see Fig.~\ref{fig:cpp}). As a consequence, in our calculations the CE phase, which is insulating in bulk at $x=0.5$, may become metallic.  We have analyzed the possibility that insulating behavior may arise at interfacial layers due to strain. We found that the interface CE layer becomes insulating (a gap opens at the Fermi energy) for relatively large values of tensile strain $e_{xy} \sim 2\%$ and $\delta=2 t$ if $\alpha \lesssim 0.7$. For larger values of $\alpha$ and/or smaller or negative values of the strain, the interface CE layer is always metallic. In reality, we expect that phase separation~\cite{brey-PRB07} or a different (maybe incommensurate with the lattice) order~\cite{calderon-nat05} may arise at the interface layers leading to insulating behavior. In this case, our results imply that an already existing insulating gap may be enhanced by tensile strain. On the other hand, compressive strain favors the occupation of the $3z^2-r^2$ orbitals which gain energy by hopping in the z-direction to the adjacent FM layer, leading to metallic behavior.

\section{Discussion and conclusions}
\label{sec:disc-concl}
It is well known that manganite thin films have properties (magnetic critical temperature $T_c$, transport) different from those exhibited by bulk manganites. In particular, it has been observed that the conductivity and the magnetic $T_c$ are reduced with respect to bulk.~\cite{infantePRB07,adamoAPL09} These differences may come about due to the effect of the strain, produced by the lattice mismatch with the substrate, throughout the whole film, and due to the modifications at the substrate/manganite interface (electronic reconstruction, phase separation), which can be most important for the very thin films used in multilayers. The electronic reconstruction may also be strongly affected by strain as it is mainly related to a redistribution of charge (although other effects, like disorder, might also play a role). We have focused here on the effect of the strain on the electronic properties of manganites at the interface with an insulator and in bulk.

The ``active'' orbitals in manganites are the Mn e$_g$: $3z^2-r^2$ and $x^2-y^2$. A tetragonal distortion of the (pseudo)cubic unit cell produces a preferred occupation of one of the anisotropic e$_g$ levels or, equivalently, to an energy splitting which is associated to a particular Jahn-Teller phonon mode and leads to orbital ordering.~\cite{tokura00,abadAFM07,nandaPRB08,sadocPRL10} In turn, a particular orbital ordering is associated to a particular magnetic ordering (see Fig.~\ref{fig:mag-conf}). Therefore, an extension of the lattice parameter in the xy-plane (tensile strain) favors the occupation of $x^2-y^2$ which is dominant in the A and CE-type AF orders while compressive strain favors $3z^2-r^2$ and, hence, the C-type order. This is exactly what we find for $x=0.5$, see Figs.~\ref{fig:bulk-lambda-JAF} and~\ref{fig:bulk-JAF-exy}, if the splitting $\delta$ is sufficiently large ($\delta \gtrsim 50 e_{xy} t$). This value of $\delta$ is relatively large ($\sim t$ for $2\%$ strain) but comparable to the Jahn-Teller splitting.~\cite{millis96} This tuning of the orbital arrangement with strain has been found in experiments,~\cite{tokura00,yamada06} model calculations for LaMnO$_3$,~\cite{nanda09} and ab-initio calculations for La$_{0.66}$Sr$_{0.33}$MnO$_3$ [\onlinecite{maJPCM07}] and for LaMnO$_3$/SrMnO$_3$ superlattices.~\cite{nandaPRB08}

The general trends are the same in the insulator/manganite ($x=0.7$)/insulator sandwich considered in Sec.~\ref{sec:interface}. The bulk behavior (FM and metallic) is recovered within a few unit cells while at the interface layer, where a charge density close to $0.5$ is expected due to the redistribution of charge, a CE order may arise. This CE order is favored by tensile strain and disfavored by compressive strain, as in bulk. 

Due to double exchange, FM and metallicity usually come hand in hand in manganites. DE is suppressed when the degeneracy of the e$_g$ levels is broken. Therefore, it is expected that uniform strain produces a reduction of the $T_c$.\cite{millisJAP98} As discussed in Sec.~\ref{sec:interface}, at an interface layer the splitting of the e$_g$ levels produced by compressive strain ($\delta>0$) is enhanced. This may also lead to a reduced conductance at interfaces with respect to bulk. 

We are assuming a uniform strain and a uniform Poisson ratio $\nu$. However, even in the case of achieving a uniform in-plane strain, the strain in the z-direction may change with the distance to an interface or surface. In this case, $\delta$ may have different signs at different atomic planes and lead to more complex orbital arrangements than the ones reported here.~\cite{sadocPRL10}

In conclusion, we have performed model calculations of the effect of uniaxial uniform strain on a bulk half-doped manganite and at the interface between a ferromagnetic metallic manganite and an insulator. The main consequence of the strain is the occurrence of an orbital selection which is intimately related to the spin degree of freedom and the transport properties. In this way, strain provides with a way to tune the ground state configuration on manganites and, therefore, control the performance of manganite based electronic devices. 

This work is supported by FIS2009-08744 (MICINN, Spain). A.B. also acknowledges the JAE program (CSIC, Spain) and  M.J.C. the Ram\'on y Cajal program (MICINN, Spain).

\bibliography{manganites}

\end{document}